\documentclass[12pt,preprint]{aastex}

\newcommand{\hst}{{\it HST}~}
\newcommand{\chandra}{{\it Chandra}~}

\newcommand{\suzaku}{{\it Suzaku}~}
\newcommand{\xmm}{{\it XMM-Newton}~}
\newcommand{\xmmn}{{\it XMM-Newton}}
\newcommand{\swift}{{\it Swift}~}
\newcommand{\erg}{{erg s$^{-1}$ cm$^{-2}$}}
\newcommand{\kms}{\ifmmode {\rm km\ s}^{-1} \else km s$^{-1}$\fi}
\newcommand{\Msun}{\ifmmode {\rm M}_{\odot} \else M$_{\odot}$\fi}
\newcommand{\Lsun}{\ifmmode {\rm L}_{\odot} \else L$_{\odot}$\fi}
\newcommand{\qo}{\ifmmode q_{\rm o} \else $q_{\rm o}$\fi}
\newcommand{\Ho}{\ifmmode H_{\rm o} \else $H_{\rm o}$\fi}
\newcommand{\ho}{\ifmmode h_{\rm o} \else $h_{\rm o}$\fi}

\newcommand{\vFWHM}{\ifmmode v_{\mbox{\tiny FWHM}} \else
                    $v_{\mbox{\tiny FWHM}}$\fi}
\newcommand{\CCF}{\ifmmode F_{\it CCF} \else $F_{\it CCF}$\fi}
\newcommand{\ACF}{\ifmmode F_{\it ACF} \else $F_{\it ACF}$\fi}
\newcommand{\halpha}{\ifmmode {\rm H}\alpha \else H$\alpha$\fi}
\newcommand{\hbeta}{\ifmmode {\rm H}\beta \else H$\beta$\fi}
\newcommand{\hgamma}{\ifmmode {\rm H}\gamma \else H$\gamma$\fi}
\newcommand{\hdelta}{\ifmmode {\rm H}\delta \else H$\delta$\fi}
\newcommand{\Lya}{\ifmmode {\rm Ly}\alpha \else Ly$\alpha$\fi}
\newcommand{\Lyb}{\ifmmode {\rm Ly}\beta \else Ly$\beta$\fi}
\newcommand{\HeI}{\ifmmode {\rm He}\,{\sc i}\,\lambda5876 \else 
	          He\,{\sc i}\,$\lambda5876$\fi}
\newcommand{\HeII}{\ifmmode {\rm He}\,{\sc ii}\,\lambda4686 \else 
	           He\,{\sc ii}\,$\lambda4686$\fi}

\newcommand{\feii}{Fe\,{\sc ii}}

\newcommand{\ciii}{\ifmmode {\rm C}\,{\sc iii} \else C\,{\sc iii}\fi}
\newcommand{\civ}{\ifmmode {\rm C}\,{\sc iv} \else C\,{\sc iv}\fi}
\newcommand{\CIV}{\ifmmode {\rm C}\,{\sc iv}\,\lambda1549 \else 
	           C\,{\sc iv}\,$\lambda1549$\fi}

\newcommand{\mgii}{Mg\,{\sc ii}}
\newcommand{\MGII}{Mg\,{\sc ii}\,$\lambda2800$}

\shorttitle{Mrk\,590 Changes Type Again}
\shortauthors{}

\received{}
\begin{document}

\title{The Changing-Look Quasar Mrk 590 is Awakening}

\author{ S.~Mathur\altaffilmark{1,2}, K.~D.~Denney\altaffilmark{3},
  A.~Gupta\altaffilmark{1,4}, M. Vestergaard\altaffilmark{5,6},
  G.~De~Rosa\altaffilmark{7}, Yair Krongold\altaffilmark{8},
  F. Nicastro\altaffilmark{9}, J. Collinson\altaffilmark{10},
  M. Goad\altaffilmark{11}, K. Korista\altaffilmark{12},
  R.~W.~Pogge\altaffilmark{1,2}, B. M. Peterson\altaffilmark{1,2,7} }

\altaffiltext{1}{Department of Astronomy, 
		The Ohio State University, 
		140 West 18th Avenue, 
		Columbus, OH 43210, USA;
		denney@astronomy.ohio-state.edu}
		
\altaffiltext{2}{Center for Cosmology and AstroParticle Physics, 
                 The Ohio State University,
		 191 West Woodruff Avenue, 
		 Columbus, OH 43210, USA}

\altaffiltext{3}{Illumination Works}

\altaffiltext{4}{Department of Biological and Physical Sciences, 
                      Columbus State Community College, 
                      Columbus, OH 43215, USA}

\altaffiltext{5}{Dark Cosmology
  Centre, Niels Bohr Institute, University of Copenhagen, Juliane Maries
  Vej 30, DK-2100 Copenhagen \O, Denmark}

\altaffiltext{6}{Steward Observatory, 933 N. Cherry Avenue, Tuscon, AZ, 85721}

\altaffiltext{7}{Space Telescope Science Institute, Baltimore, MD, USA}

\altaffiltext{8}{Instituto de Astronomia, Universidad
  Nacional Autonoma de Mexico, Cuidad de Mexico, Mexico}

\altaffiltext{9}{Istituto Nazionale di Astrofisica (INAF) – Osservatorio
  Astronomico di Roma, via Frascati, Monte Porzio Catone 00078, RM,
  Italy}

\altaffiltext{10}{Centre for Extragalactic Astronomy, Department of
  Physics, Durham University, South Road, Durham DH1 3LE, UK}

\altaffiltext{11}{Department of Physics and Astronomy, University of
  Leicester, University Road, Leicester LE1 7RH, England, UK}

\altaffiltext{12}{Department of Physics, Western Michigan University,
  Kalamazoo, MI 49008}

\begin{abstract}
  Mrk\,590 was originally classified as a Seyfert 1 galaxy, but then it
  underwent dramatic changes: the nuclear luminosity dropped by over two
  orders of magnitude and the broad emission lines all but disappeared
  from the optical spectrum. Here we present followup observations to
  the original discovery and characterization of this ``changing look''
  active galactic nucleus (AGN). The new \chandra and \hst\ observations
  from 2014 show that Mrk\,590 is awakening, changing its appearance
  again.  While the source continues to be in a low state, its soft
  excess has re-emerged, though not to the previous level. The UV
  continuum is brighter by more than a factor of two and the broad
  \mgii\ emission line is present, indicating that the ionizing
  continuum is also brightening. These observations suggest that the
  soft excess is not due to reprocessed hard X-ray emission. Instead, it
  is connected to the UV continuum through warm
  Comptonization. Variability of the Fe K-$\alpha$ emission lines
  suggests that the reprocessing region is within $\sim 10$ light years
  or $3$pc of the central source. The AGN type change is neither due to
  obscuration, nor due to one-way evolution from type-1 to type-2, as
  suggested in literature, but may be related to episodic accretion
  events.

\end{abstract}

\keywords{galaxies: active --- galaxies: nuclei --- quasars: emission lines}

%****************************************************************************
%***********MAIN BODY STARTS HERE********************************************
%****************************************************************************

%Reference Keys
%\label{Fig_junk} (just before end \figure call)
%\ref{Fig_junk}
%\label{S_junk} (just after \section call)
%\ref{S_junk}
%\label{Tab_junk} (just before end table environ)
%\ref{Tab_junk}

\section{INTRODUCTION}

Active galactic nuclei (AGNs) are characterized by continuum emission
across the electromagnetic spectrum and strong emission lines in the UV,
optical and NIR. AGNs with both broad and narrow emission lines are
classified as ``Type 1'' and those with only narrow emission lines are
``Type 2'', and this difference is usually attributed to our viewing
angle relative to an obscuring midplane. AGNs are also known to be
variable sources, showing continuum as well as emission line
variability. The continuum variability is observed on both short and
long time scales, but the variability amplitude is usually small, of the
order of $\sim 20$\% over a year for low luminosity AGNs  (i.e. with
  Seyfert-like luminosity of about $10^{41}$ to $10^{44}$ erg
  s$^{-1}$). The emission line variations usually track the continuum
variations, but see Goad et al. (2016).

%(e.g: optical: Fausnaugh et al. 2016; Aranzana et al. 2018; UV: Shapee
%et al. 2014, De Rosa et al. 2015; X-ray:Gonzalez-Martin \& Vaughan 2012,
%Edelson et al. 2015 and references therein) (e.g. Pei et al. 2016 and
%references therein)

A special class of AGN that has recently been gaining recognition has
been termed ``changing-look quasars''\footnote{We make no distinction
  among the terms quasar, AGN, and Seyfert galaxy --- all are objects
  that contain a central BH that is actively accreting sufficient matter
  to shine as a point source and that, to first order, produce the same
  emission characteristics.}.  Their broad emission lines are observed to
appear or disappear together with large changes in continuum
luminosity. The changes in the emission line spectra are such that they
change type, e.g. from ``type 1'' to ``type 2'', and this is clearly an
\underline{intrinsic} change, implying ``type'' is not always associated
with viewing angle.  This phenomenon has become of recent interest due
to a combination of several serendipitous discoveries of changing-look
AGNs (e.g. Shappee et al. 2014, Denney et al. 2014, Lamassa et al. 2015)
and the growing availability of large spectroscopic quasar databases
with long time-baselines of multi-epoch photometry and spectroscopy.
These rare occurrences have the potential to contribute to our limited
understanding of the central engines of active galaxies.  Systematic
searches of databases such as the SDSS/BOSS/TDSS have begun to uncover
more changing look quasars at larger redshifts and led to predictions
of the possible occurrence rates for quasars to change type (Runnoe et
al. 2016, MacLeod et al. 2016).

Changing-look quasars are of particular interest for their potential to
shed light on the physical origin of quasar variability and the details
of accretion, since they have undergone such an extreme apparent change.
Additionally, what is the role these objects play in the black hole (BH)
accretion history --- are we observing the beginning/end of a quasar
phase?  What role do these objects play in the larger context of galaxy
and BH co-evolution?  Recent investigations of changing-look quasars
have tried to use the observations before and after the change and any
additional data available on the few known sources to test whether or
not the changes are due to changes in accretion rate or obscuration, and
to determine if the quasar is turning on for the first time, reviving
after a period of quiescence, or something else.  In all cases so far,
obscuration has not been identified as the major cause of changing look.
Instead, a significant increase (decrease) in the mass accretion rate is
the most favored explanation for the appearance (disappearance) of the
broad emission lines.

The nearby, low-luminosity AGN  Markarian~590 (Mrk\,590 here onwards)
remains an interesting case of a changing-look quasar in the local
universe. Denney et al. (2014; D14 hereafter) have described the initial
set of observations that catalog this object's observed history and
``change'', where it went from being a strong broad-line emitter three
decades ago, allowing a reverberation mapping-based direct BH mass
measurements to be made ($(4.75 \pm 0.74) \times 10^7M_\odot$, Peterson
et al. 2004) to being broad-line weak --- the optical broad lines all
but disappeared --- in the past decade. Mrk\,590 also shows the presence
of ultra-fast outflows in the X-ray band (Gupta, Mathur \& Krongold
2015).  We have obtained additional UV and X-ray spectra since those
presented by D14 that we describe in sections 2 and 3.  Implications of
our results are discussed in section 4 and we show that the ``soft X-ray
excess'', observed in a large fraction of AGNs, is not due to the 
reprocessing of hard X-ray continuum, but is instead a result of thermal
Comptonization of UV photons. We also argue that the changing look
phenomenon is a normal event in the duty cycle of ``normal'' quasars.

\section{Chandra Spectroscopy}

Mrk~590 was observed by \chandra for $68.87$ ks on 2014 November 15
(ObsID 16109). This is longer than the $30$ ks initial DDT observation
presented by D14, but the observing parameters were otherwise kept the
same. We observed with ACIS-S/HETG; the grating was inserted in front of
the detector in order to mitigate pile-up, if any.  The data were
reduced with {\it CIAO} following the normal procedure and analyzed
using {\it XSPEC}. A Galactic column density of N$_{\rm H}=2.7 \times
10^{20}$ cm$^{-2}$ was included in all fits and was held fixed. All the
quoted errors are for $90$\% confidence, unless noted otherwise. Figure
1 shows the new, 2014 spectrum, together with the 2013 spectrum
presented by D14, and the previous ``high state'' spectrum observed in
2004. The source continues to be in a low state in 2014, so the quality
of the high resolution grating spectrum is too poor to perform useful
science. Hence we present the zeroth order spectrum here, as was done by
D14.

A simple absorbed power-law model did not provide a good fit to the 2014 
\chandra spectrum. The residuals to this fit are shown in Fig. 2, which
clearly show a soft-excess, so we added a black-body component to the
model. Clear residuals were also observed around 6 keV, so following
Longinotti et al. (2007) we added two Gaussian emission lines at
rest-frame 6.4 keV and 6.7 keV representing neutral and ionized Fe
K$\alpha$ emission lines, respectively. The resulting fit was good
($\chi^2_{\nu}=1.0$ for 104 degrees of freedom) and the parameters of
the fit are presented in Table 1 and the spectra are shown in figures 3
and 4.  Longinotti et al. (2007) required a third emission line at 7
keV; our data are not good enough at this energy range to require this
line. They also detected a line in the soft X-ray band, at $19$\AA\
(0.65 keV), but we see no evidence of that line.

The $0.3$--$10$ keV flux of the source is $2.2 \times 10^{-12}$ \erg\
($1.9 \times 10^{-12}$ \erg\ in $0.5$--$10$ keV and $2.0 \times 10^{-12}$
\erg\ in $0.3$--$8$ keV). This is significantly smaller than the 2002
flux ($8.4 \times 10^{-12}$ \erg) reported by Gallo et al. (2006),  the
2004 flux ($11 \times 10^{-12}$ \erg) reported by Longinotti et
al. (2007), and the  \suzaku flux of $6.8 \times 10^{-12}$ \erg\ in $2$--$10$
keV reported by Rivers et al. (2012).  D14  reported the low state of
Mrk~590 with $0.5$--$10$ flux of $1.1 \times 10^{-12}$ \erg, so compared
to our DDT \chandra observation in 2013, the source flux appears to have
increased slightly, but as shown in Fig. 1, the source
continues to be in a low state. The flux in the soft excess component is
$3.8 \times 10^{-13}$ \erg\ in the $0.3$--$8$ keV band (though it is
concentrated below 2 keV), which is $19\%$ of the total. If we restrict
 to the $0.3$--$2$ keV soft band, then the total flux is $9.7
\times 10^{-13}$ \erg\ while the soft excess flux remains the same as
above, making it $39\%$ of the total soft-band flux.

\section{{\it HST}/COS NUV Spectroscopy}

We obtained observations with the Cosmic Origins Spectrograph on-board
{\it HST} that were coordinated with our approved \chandra program.  The
goal of the COS observations was to observe the portion of the UV
spectrum not covered by the initial COS G140L DDT observations presented
by D14.  The new observations were obtained on 2014 November 15 with the
G230L grating and covered the observed frame wavelength ranges
$\sim$1670$-$2175\,\AA\ and $\sim$2765$-$3215\,\AA.  Importantly, this
coverage was selected to allow a continuum overlap region with the 2013
G140L spectrum and also to cover the \MGII\ doublet in this redshift
$z=0.026385$ object. The G230L spectrum was processed with the standard
COS pipelines.  Figure~\ref{F_spectra} shows the G230L COS spectrum,
that was modeled with a power-law continuum, a Balmer continuum, and an
\feii\ template (Vestergaard \& Wilkes 2001) as a means to isolate the
\mgii\ emission, with both a narrow-line component and a broad-line
component. The width and velocity shift of the narrow-line components of
\mgii\ and $\rm N III] 1750$\AA\ were tied, allowing only the strength
to differ.  We also corrected the spectrum for foreground reddening,
assuming  $E(B-V)=0.0367$mag determined from the dust maps of
Schlegel, Finkbeiner \& Davis (1998), and the extinction curve of
O'Donnell et al. (1994). As can be seen in the figure, the \MGII\
emission line clearly shows a broad component. We measured the FWHM$=
10028 \pm 627$ \kms and the line dispersion $\sigma = 4000\pm 790$ \kms\
for the broad component from the emission line models.
% Kelly's numbers: $8540 \pm 350$ \kms and the line dispersion $= 3490
% \pm 170$ \kms.

\section{Discussion}
\label{S_discuss}

\subsection{Soft excess and the Fe lines}

Over the years, X-ray observations of Mrk\,590 have shown that its
spectrum is well fit by a power-law continuum, a soft-excess, a hard
reflection component and Fe emission lines (\S 1). Rivers et al (2012),
using \suzaku data, first noticed that while the hard X-ray continuum of
the source remained practically the same as in the historic data, its
soft-excess had disappeared in 2011. In the 2013 \chandra spectrum of
D14, there was no evidence of a soft excess either, and the source
was in a historically low state.  Our new \chandra observations show
that Mrk\,590 continues to be in a low state, but a soft excess has
re-emerged. The soft-excess temperature ($kT=0.11^{+0.05}_{-0.02}$ keV)
is lower than $kT=0.18\pm 0.01$ keV seen in the 2007 \xmm spectrum
reported by Longinotti et al. (2007) and Rivers et al. (2012). The
soft-excess luminosity of $5\times 10^{40}$ \erg\ is also an order of
magnitude lower than the 2007 value of $8\pm 1 \times 10^{41}$ \erg, but
higher than the upper limit derived from the 2011 \suzaku spectra of
$3\pm 1\times 10^{40}$ \erg\ (Rivers et al. 2012). Thus the soft-excess
seems to have returned, but not to the ``normal'' level seen in the
historical high state. 

In Mathur et al. 2017 (and references therein) we discussed various
  models of the soft excess, which we briefly discuss here. The main
  contenders are (1) reflection of the hard X-ray source by the
  accretion disk (e.g., Crummy et al. 2006); (2) a warm Comptonizing
  medium around the accretion disk (e.g., Ross, Fabian, \& Mineshige
  1992; Titarchuk 1994); or (3) thermal emission from an accretion disk.
  Thermal emission from an accretion disk may appear in the soft X-ray
  band only for relatively low-mass black holes.  The re-emergence of
  the soft excess in the low state, when the the hard X-ray emission was
  weak, suggests that the soft excess is not due to reprocessed hard
  X-ray emission. An alternative model, in which the soft excess is
  generated by thermal Comptonization in the optically thick accretion
  disk corona, is a more likely explanation. 

Using a series of observations of Mrk 509, Mehdipour et al (2011)
found that the soft excess is correlated with the UV flux as $L_{\rm
  soft excess}=1.261 (F_{\rm UVW2})^{2.869}$, where $L_{\rm soft
  excess}$ is the the 0.3--2 keV soft-excess luminosity in units of
$10^{42}$ erg s$^{-1}$  and $F_{\rm UVW2}$ is the flux in the UVW2 filter in
UVOT on \swift and OM on \xmmn, centered at 2030\AA, in units of
$10^{-14}$ \erg $\AA^{-1}$. In our new observations, the soft excess
luminosity of Mrk\,590 in the 0.3--2 keV band is $3.5 \times 10^{40}$
  erg s$^{-1}$ and the continuum flux at 2030\AA\ is $7.7\times 10^{-16}$
\erg $\AA^{-1}$.  The observed soft excess luminosity is two orders of
magnitude larger that that predicted by the Mehdipour relation
(correcting the UV flux for the distance of NGC\,5548) and shows
that the Mehdipour relation between the soft excess and the UV flux is
not universal.

Compared to the values in 2004 reported by Longinotti et al (2007), the
Fe $K\alpha$ emission lines have smaller equivalent widths and smaller
fluxes. This shows that both the neutral and the ionized lines have
responded to the continuum changes between 2004 and 2014. This places
the reprocessing material within 10 light-years or 3 pc of the continuum
source, and not from the kiloparsec-scale extended emission region as
suggested by Longinotti et al (2007) for the ionized line.

\subsection{The \mgii\ emission line}

Our \hst spectrum clearly shows a broad \mgii\ emission line. Since we
do not have a UV spectrum covering the \mgii\ line in the low state when
the \hbeta\ line had vanished, the presence of the \mgii\ line in the
current spectrum can be interpreted in a couple of different ways.

Photoionization models and evidence from reverberation mapping of the
ionization stratification of the BLR support the co-spatial existence of
\hbeta\ and \mgii\ emitting gas. Thus it is possible that when the broad
\hbeta\ emission line disappeared, the broad \mgii\ line had also
disappeared, and now that a broad \mgii\ line has emerged, possibly a
broad \hbeta\ has also emerged.  We should note, however, that the
response of a line to continuum variations depends not only on the
geometry and location of line emitting region, but also on line
responsivity. Cackett et al. (2015) have shown that the \mgii\ emission
line has very low responsivity compared to other high-ionization
emission lines (see also Korista \& Goad 2000). It is thus possible that
the broad \mgii\ line had never disappeared. If this is the case, then we
cannot comment on the presence of a broad \hbeta\ line in 2014.  Roig et
al. (2014) found a class of AGNs in the Baryon Oscillation Spectroscopic
Survey showing a broad \mgii\ without a broad
\hbeta. Wavelength-dependent scattering of the broad-line region is
offered as an explanation by these authors as to why the broad Balmer
lines could be missing but the broad \mgii\ lines are strong. Our
results on Mrk\,590 suggest that such AGNs may be in a transition state
when \hbeta\ disappeared, but \mgii\ did not, owing to its low
responsivity.

The connection between the soft excess and the emission lines is of
interest. In the warm Comptonization model, the seed UV/EUV photons are
Compton up-scattered into soft X-rays in the optically thick accretion
disk corona. Observationally, the soft-excess has been found to
correlate with UV emission (e.g., Atlee \& Mathur 2009). Thus the
emergence of soft excess in Mrk\,590 implies a corresponding emergence
of the EUV ionizing continuum. This in turn may be responsible for the
re-appearance of the broad-line region, which we see with \mgii. This is
our preferred model. D14 found the continuum flux at 1450\AA\ to be
$3.7\times 10^{-16}$ \erg \AA$^{-1}$ which was two orders of magnitude
lower than the {\it IUE} value (D14). The 1634\AA\ flux in our new \hst
data is $9.2\times 10^{-16}$ \erg \AA$^{-1}$, which extrapolated to
1460\AA\ is $15.1\times 10^{-16}$ \erg \AA$^{-1}$. Thus, while the
source continues to be in a low state, the UV continuum is re-emerging
along with the soft-excess. While the presence of the broad \mgii\
  line can be interpreted in different ways, as noted above, it is
  consistent with the increasing luminosity of the ionizing continuum.

%The evidence from \mgii\ observations suggests
%that the ionizing continuum is also increasing in luminosity.

\section{Conclusion}
\label{S_conclude}

D14 presented optical, UV, and X-ray observations of the classical
Seyfert 1 Mrk\,590 that span the past 40+ years.  This interesting
object brightened by a factor of a few tens between the 1970s and 1990s
and then faded by a factor of a 100 or more at all continuum wavelengths
between the mid-1990s and 2013.  Notably, there is no evidence in the
current data set that this recent, significant decline in flux is due to
obscuration; in particular, the most recent X-ray observations are
consistent with zero intrinsic absorption.  There were similarly
dramatic changes in the emission-line fluxes, the most striking of which
is the complete disappearance of the broad component of the \hbeta\
emission line, which had previously been strong (equivalent widths
$\sim$20$-$60 \AA).

In this paper we have presented new \chandra and \hst\ observations from
2014 and we find that Mrk\,590 is awakening. While the source continues
to be in a low state, its soft excess has re-emerged, though not to the 
historical level. The broad \mgii\ emission line is also present. These
observations suggest that the soft excess is connected to the UV
continuum through warm Comptonization and that the ionizing continuum is
also on the rise, but it does not follow the relation between soft
excess and UV continuum reported in Mehdipour et al. (2011).

The implications from this long time series of Mrk\, 590
observations are that (1) Mrk\,590 is a direct challenge to the
historical paradigm that AGN type is exclusively a geometrical effect,
and (2) there may not be a strict, one-way evolution from Type 1 to Type
1.5$-$1.9 to Type 2 as recently suggested by Elitzur, Ho, \& Trump (2014).
Instead, for at least some objects, the presence of BLR emission may
coincide  with episodic accretion events throughout a single active
phase of an AGN.  If true, such behavior may be more prominent in
Seyfert galaxies, where accretion is likely to be a consequence of
secular processes (e.g., Mathur et al. 2012; Martin et al. 2018) and
therefore likely more episodic than quasar activity, which may be
governed more predominantly by major mergers.

%The changing look AGNs like Mrk\,590, in which the broad \hbeta\ line
%disappeared, provide a cautionary note to using the line widths to
%measure black hole masses.

\acknowledgements Support for this work was provided by the National
Aeronautics and Space Administration through Chandra Award Number
G04-15114X to SM issued by the Chandra X-ray Observatory Center, which
is operated by the Smithsonian Astrophysical Observatory for and on
behalf of the National Aeronautics Space Administration under contract
NAS8-03060. Support for the \hst program number GO-13185 was provided by
NASA through a grant from the Space Telescope Science Institute, which
is operated by the Association of Universities for Research in
Astronomy, Inc., under NASA contract NAS5-26555. KDD was supported by an
NSF AAPF fellowship awarded under NSF grant AST-1302093 while she worked
on this project. MV gratefully acknowledges support from the Danish
Council for Independent Research via grant number DFF 4002-00275.

%****************************************************************************
%***********BIBLIOGRAPHY STARTS HERE*****************************************
%****************************************************************************

%****************************************************************************
%***********TABLES START HERE************************************************
%****************************************************************************

\begin{table}
%\scriptsize
\caption{\chandra best-fit spectral model$^1$ }
\begin{tabular}{lcccc}
\hline
Model  &  Parameter 1  &  Normalization$^2$  & Parameter 3  \\
        &          &    &      \\
\hline
 Power-law &  $\Gamma=1.6\pm 0.1$&  $2.4 \pm 0.3\times 10^{-4} $  &    \\
 Black Body &  kT$=0.11^{+0.05}_{-0.02}$ keV &  $7^{+7}_{-4} \times 10^{-6}$  &        \\
Line 1& E $=6.4$ keV$^{\dagger}$ & $5 \pm 2 \times 10^{-6}$ & EW $= 268\pm 100$ eV   \\
Line 2& E $=6.7$ keV$^{\dagger}$ &  $4 \pm 2 \times 10^{-6}$ &  EW $=176\pm 100$ eV  \\
&&&    \\
\hline
\end{tabular}

\noindent
1. $\chi^{2}_{\nu}$ for the joint fit is 1.0 for $\nu=104$ dof. \\
2. in units of photons s$^{-1}$ cm$^{-2}$ keV$^{-1}$ \\
$\dagger$: Rest-frame; parameter frozen.\\
\end{table}

\begin{figure*}
%\figurenum{1}
\epsscale{0.5}
\plotone{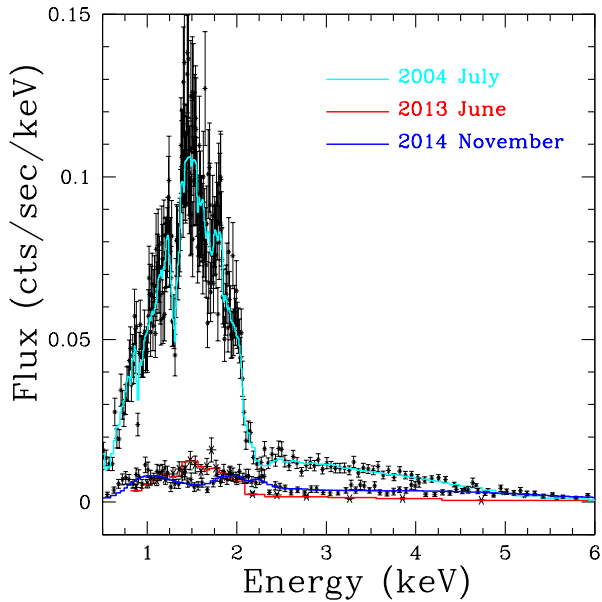}
%\plotone{f1.eps}
\caption{{\it Chandra} spectra of Mrk\,590. The high state spectrum of
  2004 and the low state spectra of 2013 and 2014 are shown. The spectra
  are fitted with a simple power-law continuum models modified by
  Galactic absorption; these are shown with colored lines to guide the
  eye. The purpose of this figure is to show that the source continues
  to remain in the low state in 2014. }
%\label{F_xrayspec}
\end{figure*}

\begin{figure*}
%\figurenum{1}
%\epsscale{0.5}
\includegraphics[height=3in,angle=-90]{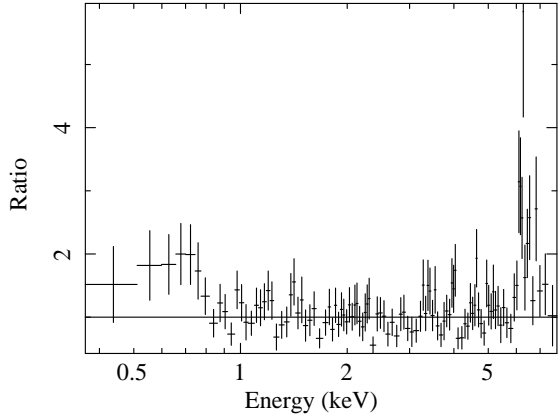}
%\plotone{ratio.ps,angle=-90}
%\plotone{f1.eps}
\caption{ Residuals to the 2014 spectrum fitted with an absorbed
  power-law model. The soft excess is clearly visible below 1 keV. Note
  also the residuals between 6 keV and 7 keV, at the locations of Fe
  lines.}
%\label{F_xrayspec}
\end{figure*} 

\begin{figure*}
%\figurenum{1}
%\epsscale{0.5}
\includegraphics[height=3in,angle=-90]{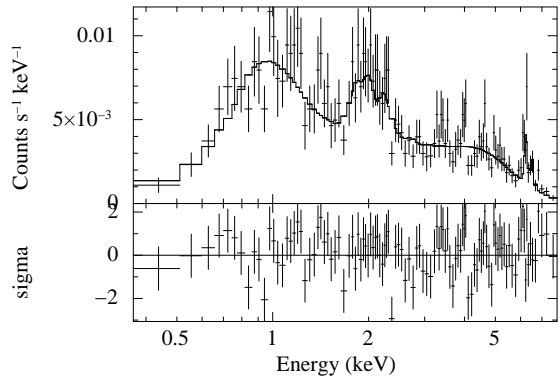}
%\plotone{pl+2gauss+bbfit.ps,angle=-90}
%\plotone{f1.eps}
\caption{The best fit spectrum and the residuals. The spectrum is fitted
  with an absorbed power-law, a black-body component to parameterize the
  soft excess and two emission lines at 6.4 keV and 6.7 keV (rest
  frame). }
%\label{F_xrayspec}
\end{figure*} 

\begin{figure*}
%\figurenum{1}
%\epsscale{0.5}
\includegraphics[height=3in,angle=-90]{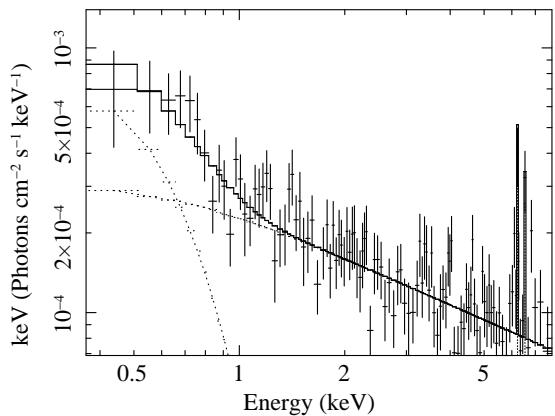}
%\plotone{eufspec.ps,angle=-90}
%\plotone{f1.eps}
\caption{The ``unfolded'' spectrum, i.e. the spectrum without the
  instrumental response folded in. The model components (power-law
  continuum, soft excess, and emission lines) are shown as dotted
  lines. }
%\label{F_xrayspec}
\end{figure*} 

\begin{figure*}
%\figurenum{2}
\epsscale{1.0}
\includegraphics[height=5in,angle=90]{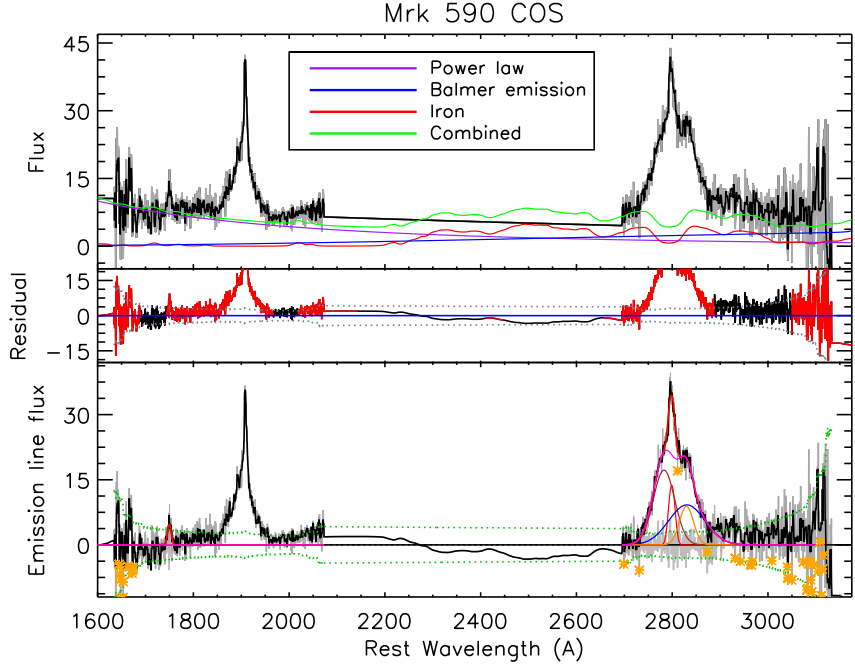}
%\plotone{f2.eps}
\caption{ Milky Way extinction corrected HST/COS spectrum of Mrk 590
  with continuum and line models. Shown are observed flux densities
  shown as function of the restframe wavelengths. {\em Top panel:} The
  observed spectrum with the power-law continuum model (purple curve),
  the Balmer continuum model (blue curve), the iron emission model (red
  curve), and the sum of the three models (green curve) shown
  superimposed. {\em Middle panel:} Residuals obtained by subtracting
  the continuum, iron, and Balmer emission models shown along with the
  flux errors (gray dotted curve). Indicated in black are the wavelength
  regions used to fit the power-law continuum. The \mgii\ emission line
  is clearly broad. {\em Bottom panel:} Emission line spectrum (black)
  obtained by subtracting the continuum, iron, and Balmer emission
  models (shown in the top panel) from the observed spectrum with the
  line models superimposed.  Model-fit were made to NIII]1750\AA{} (red
  curve) and the broad (magenta curve) and narrow (red curves)
  components of \mgii. Red curves show the narrow line components while
  brown, blue and yellow curves represent the three Gaussian functions
  used to model the broad \mgii\ component (magenta curve). The gray
  curve (around zero flux level) shows the residuals after subtracting
  the emission line model. Yellow stars indicate pixels deviating more
  than 3$\sigma$ from the rms in the spectrum, which were filtered out
  prior to the line modeling. The flux errors are shown by the green
  dotted curve.  }
\label{F_spectra}
\end{figure*}


\begin{thebibliography}{}

\bibitem[]{}Aranzana, E., Kording, E., Uttley, P., Scaringi, S. \&
  Bloemen, S. 2018, MNRAS, 476, 2501

\bibitem[Arnaud et al. 1985]{}Arnaud, K. et al. 1985, MNRAS, 217, 105

\bibitem[Atlee \& Mathur 2009]{}Atlee, D. \& Mathur, S. 2009, ApJ, 703, 1597

\bibitem[Bianchi et al. 2009]{}Bianchi et al. 2009, A\&A, 495, 421

\bibitem[]{}Cackett, E.M., Gultekin, M., Bentz, M. et al. 2015, ApJ, 810, 86

\bibitem[Crummy et al. 2006]{}Crummy, J., Fabian, A. C., Gallo, L., \& Ross, 
  R. R. 2006, MNRAS, 365, 1067

\bibitem[Denney et al. 2014]{}Denney, K., De Rosa, G., Croxall, K. et
  al. 2014, ApJ, 796, 134

\bibitem[Edelson et al. 2015]{}Edelson, R. et al. 2015, ApJ, 806,
  129. (Paper II)

\bibitem[]{}Elitzur, M., Ho, L., \& Trump. J., 2014, MNRAS, 438, 3340

\bibitem[Gallo et al. 2006]{}Gallo, L., Lehmann, I., Pietsch, W. et
  al. 2006, MNRAS, 365, 688

\bibitem[]{}Gupta. A., Mathur, S. \& Krongold, Y. 2015, ApJ, 798, 4

\bibitem[]{}Korista, K. \& Goad, M., 2000, ApJ, 536, 284

\bibitem[Longinotti et al. 2007]{}Longinotti, A., Bianchi, S.,
  Santos-Lieo, M.  et al. 2007, A\&A, 470, 73

\bibitem[]{}LaMassa, S., Cales, S., Moran, E. et al. 2015, ApJ, 800, 144

\bibitem[]{}MacLeod, C., Ross, N., Lawrence, A. et al. 2016, 457, 389

\bibitem[]{}Martin, G., Kaviraj, S., Volonteri, M. et al. 2018, MNRAS,
  476, 2801

\bibitem[]{}Mathur, S.; Fields, D.; Peterson, B. M.; \& Grupe, D., 2012,
  ApJ, 754, 146

\bibitem[Mathur et al. 2017]{}Mathur, S., Gupta, A., Page, K. et
  al. 2017, ApJ, 846, 55

\bibitem[ Mehdipour et al. 2011]{}Mehdipour, M. et al. 2011, A\&A, 534,39

\bibitem[O'Donnell]{}O'Donnell, J. E. 1994, ApJ, 422, 158O

\bibitem[Page et al. 2004]{}Page, K. et al. 2004, ApJ, MNRAS, 349, 57

\bibitem[]{}Peterson, B.M., Ferrarese, L., Gilbert, K. et al. 2004, ApJ,
  613, 682

\bibitem[Rivers 2012]{}Rivers, E., Markowitz, A., Duro, R., \&
  Rothschild, R., 2012, ApJ, 759, 63

\bibitem[Roig 2014]{}Roig, B., Blanton, M., \& Ross, N., 2014, ApJ, 781, 72

\bibitem[Ross et al. 1992]{}Ross, R. R., Fabian, A. C., \& Mineshige, S.
 1992, MNRAS, 258, 189

\bibitem[]{}Runnoe, J., Cales, S., Ruan, J., et al. 2016, MNRAS, 455, 1691

\bibitem[]{}Schlegel, D., Finkbeiner, D., \& Davis, M., 1998, ApJ, 500, 525

\bibitem[]{}Shapee, B., Prieto, J., Grupe, D. et al. 2014, ApJ, 788, 48

\bibitem[]{}Titarchuk, L., 1994, ApJ, 434, 313

\bibitem[Vestergaard \& Wilkes 2001]{}Vestergaard, M. \& Wilkes,
  B. 2001, ApJS, 134, 1

\bibitem[]{}

\end{thebibliography}
\end{document}